# Giant orbital-magnon conversion driven perpendicular magnetization switching

Fanyu Meng[1], Ying Feng[1], Mingyang Sun[1], Baiyan Kang[4], Donglin Song[1], Tuo Zhang[1], Jia Zhang[5], Wenping Zhou[4], Jijun Zhao[2,3]*, Yi Wang[1]*

*[1]Key Laboratory of Materials Modification by Laser, Ion and Electron Beams (Ministry of Education), School of Physics, Dalian University of Technology, Dalian 116024, China*
*[2]Guangdong Basic Research Center of Excellence for Structure and Fundamental Interactions of Matter, Guangdong Provincial Key Laboratory of Quantum Engineering and Quantum Materials, School of Physics, South China Normal University, Guangzhou 510006, China*
*[3]Guangdong-Hong Kong Joint Laboratory of Quantum Matter, Frontier Research Institute for Physics, South China Normal University, Guangzhou 510006, China*
*[4]Inner Mongolia Key Laboratory of Microscale Physics and Atom Innovation, and Research Center for Quantum Physics and Technologies, Inner Mongolia University, Hohhot 010021, China*
*[5]School of Physics and Wuhan National High Magnetic Field Center, Huazhong University of Science and Technology, Wuhan 430074, China*
[*]Correspondence to yiwang@dlut.edu.cn, zhaojj@scnu.edu.cn

**Abstract**

The pursuit of beyond-Moore information technologies has stimulated the exploration of novel information carriers, such as electron spin, orbital, and magnon, beyond electron charge. Efficient interconversion among these degrees of freedom and precise control over the information states are crucial for advancing nanoelectronic devices. However, a direct coupling between orbital angular momentum ($L$) and magnons ($M$) has remained elusive, and magnetization switching through orbital-to-magnon ($L$-$M$) conversion has not yet been achieved. Here, we report the experimental demonstration of $L$-$M$ conversion in an orbital metal/antiferromagnetic insulator bilayer at room temperature, with an efficiency over an order of magnitude higher than that in traditional orbital systems lacking the *L*-*M* process. Consequently, we achieved efficient room-temperature perpendicular magnetization switching in a CoFeB ferromagnetic layer mediated by this new mechanism. Our findings establish a direct link between orbitronics and magnonics, providing a new platform for the development of advanced nano-devices based on orbital-driven magnonic phenomena.

**Introduction**

Beyond the electron charge (*C*), information can be carried by various degrees of freedom, such as electron spin (*S*), orbital (*L*), and magnon (*M*). Building on traditional electronics that rely on electron charge, emerging fields like spintronics, orbitronics, and magnonics primarily focus on electron spin, orbital, and magnon, respectively. These innovative research areas hold potential for next-generation nanoelectronic devices for data processing and storage, offering better performance than conventional electronics[1-8]. However, the main challenges in realizing this potential include efficient conversion between different degrees of freedom and robust switching of binary information states within devices, as illustrated in Fig. 1.

Historically, efficient charge-to-spin (*C*-*S*) conversion (Fig. 1a) has played a vital role in spintronics, leading to the discovery of two milestone effects: spin-transfer torque (STT)[9-11] and spin-orbital torque (SOT)[2,12,13]. This has opened the era for electrically controlled spintronic device applications, such as STT- or SOT-based magnetic random access memories (MRAMs). Meanwhile, electron spin-to-magnon (*S*-*M*) conversion (Figs. 1a and 1b) is a key mechanism for effectively generating magnon currents[3,4]. Magnons enable heat-free and low-damping transfer of spin information over long distances in magnetic insulators, thus avoiding Joule heating caused by charge flow[14-16]. Magnetization switching driven by magnon current-induced magnon torque has been achieved in antiferromagnetic devices[17-22] (Fig. 1b). Nevertheless, the efficiency of magnon current excitation still needs further improvement.

Recently, orbital angular momentum (*L*) has emerged as an important degree of freedom in spintronics, due to the capability of orbital currents to generate sizable orbital torque (OT). This offers a promising pathway toward efficient magnetization switching via sequential charge-to-orbital (*C*-*L*) and orbital-to-electron spin (*L*-*S*) conversions[23-25] (Fig. 1a). The magnitude of OT is expected to be 10-100 times higher than that of SOT[26-29]. Furthermore, orbital currents can be generated in a wide range of materials with weak or negligible spin-orbital coupling, such as Ti, Ru, and Cr[27,28,30], thereby expanding the material palette far beyond traditional SOT-based platforms. However, a major challenge in the practical application of OT is the inherently low *L*-*S* conversion coefficient in common ferromagnets, such as CoFeB and NiFe, which are widely used in commercial magnetic devices. This inefficiency severely limits the OT strength in orbital material-ferromagnet heterostructures and hinders perpendicular magnetization switching[26,31-33].

These recent advances clearly highlight the importance of conversion between different degrees of freedom. It is of great interest to ask whether orbitals can effectively excite magnons and thereby mediate efficient magnetization switching, as depicted in Fig. 1c. This would not only represent an intriguing physical process but also connect the fields of orbitronics and magnonics. In this work, we successfully achieve the first perpendicular magnetization switching driven by a giant *L*-*M* conversion at room temperature using an orbital metal Ti/antiferromagnetic insulator NiO bilayer. In this process, a strong magnon current can be excited by orbital current with an impressive *L*-*M* conversion coefficient. Our work unlocks novel materials and a powerful mechanism for magnetization control, and provides new insights for advancing both orbitronics and magnonics.

## Results

### Observation of orbital-to-magnon conversion and orbital-induced magnon torque

To investigate *L*-*M* conversion, we fabricate devices with the structure of Ti (5 nm)/NiO ($t_{NiO}$)/ferromagnet (FM, 6 nm), where FM represents CoFeB ($Co_{20}Fe_{60}B_{20}$) or Ni (Fig. 2a). Ti is chosen as the orbital-Hall material due to its negligible spin-orbital coupling and large orbital Hall conductivity (~$7.6\times10^3$ $\hbar/2e$ $\Omega^{-1}cm^{-1}$)[5,29,34]. An orbital current is generated by the charge current flowing through the Ti layer, and is injected into the adjacent NiO layer. The damping-like torque efficiency ($\theta_{DL}$) is characterized using the spin-torque ferromagnetic resonance (ST-FMR) technique[12,35,36]. Figure 2b shows a representative ST-FMR signal ($V_{mix}$) for the Ti (5 nm)/NiO (3 nm)/CoFeB (6 nm) device at 6 GHz, and the fits. By adopting the established analysis method[37,38], we evaluate the damping-like torque $\theta_{DL}$ from the symmetric component. All ST-FMR measurements in this study are conducted at room temperature.

Figure 2c shows the measured $\theta_{DL}$ as a function of NiO thickness for Ti (5)/NiO ($t_{NiO}$)/CoFeB (6) and Ti (5)/NiO ($t_{NiO}$)/Ni (6) devices (unit in nm). When $t_{NiO}$ = 0 nm, the $\theta_{DL}$ of Ti/Ni bilayer is much larger than that of Ti/CoFeB bilayer due to the larger *L-S* conversion coefficient in Ni[5,39]. As $t_{NiO}$ increases, $\theta_{DL}$ rises continuously in the Ti/NiO/CoFeB device, indicating the emergence of a magnon torque alongside the direct orbital torque, since magnons are the sole carriers of spin angular momentum in the insulating NiO layer[14,15,17]. In the thin NiO regime ($t_{NiO} < 3$ nm), $\theta_{DL}$ clearly depends on the ferromagnet, suggesting that orbital currents can still partially penetrate the thin NiO layer and generate different orbital torques in CoFeB versus Ni. Notably, at $t_{NiO} = 3$ nm, $\theta_{DL}$ peaks for both devices, reflecting the coexistence of orbital and magnon torques. For $t_{NiO} > 3$ nm, $\theta_{DL}$ decreases gradually for both devices due to magnon-current attenuation in thicker NiO. Finally, at $t_{NiO} = 10$ nm, the values of $\theta_{DL}$ for CoFeB and Ni converge, confirming that only the magnon torque remains. This result also rules out the possible contributions from the differences in the NiO/FM interfaces. The detailed mechanism for the NiO thickness-dependent $\theta_{DL}$ is discussed later.

To further verify the *L-M* conversion and the orbital-induced magnon torque, we fabricate Hall-bar devices of Ti (5)/NiO (3 and 5)/Ti (2)/CoFeB (0.9)/MgO (2)/$SiO_2$ (4) (unit in nm), which exhibit perpendicular magnetic anisotropy (PMA). The $\theta_{DL}$ is characterized by using an independent technique of second-harmonic Hall voltage measurements[40,41], as shown in Fig. 2d. The measurement procedures are detailed in Supplementary Note 1 (also see Supplementary Figs. 1 and 2). Figure 2e shows the temperature-dependent $\theta_{DL}$ for devices with $t_{NiO}$ = 3 and 5 nm, respectively. The $\theta_{DL}$ exhibits a pronounced peak at a characteristic temperature for each device (i.e., ~60 K for the device with 3 nm NiO and ~80 K for the device with 5 nm NiO). These temperatures coincide with the respective magnetic ordering temperatures of the NiO layers, as independently determined from the exchange-bias measurements on the stacks of Ti (5)/NiO (3 and 5)/NiFe (3)/$SiO_2$ (4) (unit in nm), as shown in Fig. 2f. It is known that magnon currents can be significantly boosted due to the enhanced spin fluctuations in antiferromagnets near the Néel temperature[17,42-44]. Therefore, our observed behavior of $\theta_{DL}$ provides strong evidence for the magnon-originated torque and the orbital-to-magnon conversion process.

**Orbital-to-magnon conversion coefficient in NiO**

The *L*-*M* conversion coefficient is a key physical parameter that connects orbitronics and magnonics. Therefore, quantifying this coefficient in magnetic insulators is essential for practical applications. Figure 3a illustrates the *L*-*M* conversion process at the interface between an orbital material (e.g., Ti) and a magnetic insulator (MI, e.g., NiO). The total orbital current ($J_O$) generated in the Ti layer splits into two paths ($J_O = J_{O1}+J_{O2}$). In one path, the orbital current $J_{O1}$ converts into a magnon current near the interface, which then diffuses through MI and enters the FM layer. In the other path, the remaining orbital current, $J_{O2}$, directly diffuses through MI and subsequently enters the FM layer. For our devices, $\sigma_O$ is the orbital Hall conductivity of Ti, and $\sigma_O^T\big|_{t_{NiO}=0}$ is the orbital Hall conductivity at the Ti/NiO interface, associated with the orbital current $J_{O2}$. Therefore $\sigma_O - \sigma_O^T\big|_{t_{NiO}=0}$ describes the contribution of the orbital current $J_{O1}$ at the Ti/NiO interface. $\sigma_O^{eff} (= C_{FM}\sigma_O^T)$ is the effective spin Hall conductivity reflecting the spin current converted from the partial orbital current injected into the FM layer (characterized by the orbital Hall conductivity $\sigma_{\mathrm{O}}^{\mathrm{T}}$), and $C_{\mathrm{FM}}$ is the *L-S* conversion coefficient in FM. Meanwhile, $\sigma_{\mathrm{M}}^{\mathrm{eff}}[=C_{\mathrm{MI}}(\sigma_{\mathrm{O}}-\sigma_{\mathrm{O}}^{\mathrm{T}}\big|_{t_{\mathrm{NiO}}=0})]$ is the magnon-originated effective Hall conductivity that characterizes the magnon current injected into the FM layer due to the *L-M* conversion, and $C_{\mathrm{MI}}$ is the *L-M* conversion coefficient in the NiO. Therefore, the total measured Hall conductivity for the Ti/NiO/FM devices can be written as $\sigma_{\mathrm{Ti/NiO/FM}}=\sigma_{\mathrm{M}}^{\mathrm{eff}}+\sigma_{\mathrm{O}}^{\mathrm{eff}}$ [32,39,45]. The $\sigma_{\mathrm{Ti/NiO/FM}}$ can be experimentally evaluated by $\sigma_{\mathrm{Ti/NiO/FM}} = (\hbar/2e)(\theta_{\mathrm{DL}}/\rho_{\mathrm{Ti}})$ from the ST-FMR measurement data in Fig. 2c.

Based on this model, both $\sigma_{\mathrm{M}}^{\mathrm{eff}}$ and $\sigma_{\mathrm{O}}^{\mathrm{eff}}$ can be extracted for all the devices with varying NiO thicknesses, as plotted in Fig. 3b (see detailed calculation process in Supplementary Note 2). As $t_{\mathrm{NiO}}$ increases, $\sigma_{\mathrm{M}}^{\mathrm{eff}}$ first increases ($t_{\mathrm{NiO}} < 3$ nm), then peaks at $t_{\mathrm{NiO}} = 3$ nm, and finally decreases slowly. This indicates that the NiO film quality and *L-M* conversion improve as $t_{\mathrm{NiO}}$ increases. Whereas $\sigma_{\mathrm{O}}^{\mathrm{eff}}$ decreases monotonically and finally becomes negligible at $t_{\mathrm{NiO}} = 10$ nm, indicating that the orbital current is almost completely blocked by a thicker NiO. Note that orbital-induced magnon torque dominates over orbital torque as $t_{\mathrm{NiO}}$ exceeds 2 nm. As shown in Fig. 3b, by fitting $\sigma_{\mathrm{O}}^{\mathrm{eff}}$ vs. $t_{\mathrm{NiO}}$ with equation of $\sigma_{\mathrm{O}}^{\mathrm{eff}}=C_{\mathrm{FM}}\,\sigma_{\mathrm{O}}^{\mathrm{T}}\big|_{t_{\mathrm{NiO}}=0}\cdot e^{-t_{\mathrm{NiO}}/\lambda}$, we can determine the orbital diffusion length $\lambda_{\mathrm{O}} = 3.182$ nm in NiO. This result indicates that the *L-M* conversion occurs mainly near the Ti/NiO interface. Meanwhile, the magnon diffusion length $\lambda_{\mathrm{M}}$ in NiO is determined to be 25.4 nm, which is consistent with previous reports[17,22] (see Supplementary Note 2 and Supplementary Fig. 3).

Consequently, the *L-M* conversion coefficient in NiO can be quantitatively determined by $C_{\mathrm{NiO}}=\sigma_{\mathrm{M}}^{\mathrm{eff}}/(\sigma_{\mathrm{O}}-\sigma_{\mathrm{O}}^{\mathrm{T}}\big|_{t_{\mathrm{NiO}}=0})$ and the value is ~0.033 (Supplementary Note 3). Furthermore, the

orbital Hall conductivity of 5-nm Ti is evaluated to be ~$4.3\times10^{3}$ $\hbar/2e$ $\Omega^{-1}cm^{-1}$, which is consistent with previous reports[39,46]. When we consider the NiO/FM interface transparency $T_{int}$ of 0.567 from ST-FMR linewidth analysis[46,47], the final *L-M* conversion coefficient is ~0.06 (Supplementary Note 3). It is worth noting that this is a lower bound because magnon-current attenuation in NiO has not been included. Our results demonstrate that, beyond the conductive materials previously studied, magnetic insulators can convert orbital angular momentum into magnons with remarkable efficiency, as shown in Fig. 3c. This will greatly expand material options for orbitronic research.

**Orbital-induced magnon torque-driven perpendicular magnetization switching**

We demonstrate robust *L-M* conversion-driven perpendicular magnetization switching in CoFeB at room temperature, a material widely used in commercial magnetic memory. Figure 4a shows the device structure and the switching measurement configuration. The CoFeB layer exhibits good PMA, as confirmed by the square hysteresis loop of the anomalous Hall resistance ($R_{xy}$) driven by the magnetic field along the *z*-axis (Fig. 4b, blue curve). Upon sweeping pulsed current with a fixed width of 50 μs under an in-plane external magnetic field $H_x$ of +50 Oe, a distinct perpendicular magnetization switching in CoFeB is observed at either ~ +10 or ~ –10 mA (green). It shows a nearly 100% switching ratio. However, as we reverse the direction of $H_x$ (i.e., –50 Oe), the switching polarity also reverses (red), which is the characteristic of current-induced torque-driven magnetization switching.

Control experiments further confirm the crucial roles of both the orbital material (Ti) and the antiferromagnetic insulator (NiO) in magnetization switching. No magnetization switching occurs in NiO (3)/Ti (2)/CoFeB or Ti (2 and 5)/CoFeB devices (unit in nm) when the bottom Ti layer or the Ti/NiO bilayer is removed (Supplementary Note 4 and Supplementary Figs. 4a-4c). Additionally, switching is not observed in Ti (5)/MgO (3)/Ti (2)/CoFeB devices (unit in nm), where NiO is replaced with a nonmagnetic insulator MgO (Supplementary Note 4 and Supplementary Fig. 4d). These results strongly highlight the synergistic effects of Ti/NiO bilayer, supporting the orbital-induced magnon torque mechanism. Additionally, we also ruled out the Ti and NiO interface contribution in our sample (see Supplementary Note 5 and Supplementary Fig. 5).

Figures 4c-4e present the magnetization switching images captured at room temperature using a magneto-optical Kerr effect (MOKE) microscope. The device is patterned into a 40-μm-long rectangular channel. The magnetization is first initialized along the +*z* direction, corresponding to the light contrast in Fig. 4c (red box). Subsequently, the magnetization switching is induced by injecting a pulsed current $I_{pulse}$ either parallel or antiparallel to the in-plane external magnetic field $H_x$, denoted by the dark and light contrast in Figs. 4d and 4e, respectively. This switching behavior is repeatable in devices with a thicker NiO layer (Supplementary Note 6 and Supplementary Fig. 6).

Figure 4f shows the current-induced magnetization switching under $H_x$ of +50 Oe across temperatures from 5 K to 300 K in the same device. Using the extracted critical switching current density $J_c$ from Fig. 4f and the independently measured depinning field $H_P$, the magnetization switching efficiencies $\eta = H_P/J_c$ are evaluated at different temperatures[48,49]. As shown in Fig. 4g, the $\eta$ peaks near 60 K, matching the temperature at which the damping-like torque is maximum in the second-harmonic measurements (Fig. 2e).

**Superior performance of *L-M* conversion in magnetization switching**

We demonstrate that *L-M* conversion enables more efficient magnetization switching compared to conventional orbital-only or spin-only torque schemes. Firstly, the torque efficiency in Ti/NiO/CoFeB increases by more than 20 times compared to the Ti/CoFeB traditional orbital bilayer that lacks *L-M* conversion (Fig. 3b). Furthermore, we design devices of NM (5)/NiO ($t_{NiO}$ = 0, 1.5, 2, 3, 5, and 7)/Ti/CoFeB (unit in nm) with PMA, where the normal metal (NM) is W or Ta. Figures 5a and 5b display the current-induced perpendicular magnetization switching curves in W-based devices under in-plane magnetic fields $H_x$ = +50 Oe (Fig. 5a) and $H_x$ = −50 Oe (Fig. 5b). Notably, when $t_{NiO} \leq 2$ nm, the switching polarity indicates a negative torque efficiency in the devices, consistent with the negative spin Hall angle of W. When $t_{NiO} \geq 3$ nm, however, the switching polarity reverses (denoted by the blue and red arrows), indicating a change in the dominant torque direction. This behavior is corroborated by the MOKE imaging measurements (Supplementary Note 7 and Supplementary Figs. 7-9) and is also observed in Ta-based devices (Supplementary Note 7 and Supplementary Fig. 10).

Similar to Figs. 2c and 3b, the NiO thickness dependence of $\theta_{DL}$ (Fig. 5c), $\sigma_M^{eff}$ and $\sigma_O^{eff}$ (Fig. 5d) in W (5)/NiO ($t_{NiO}$)/FM (CoFeB or Ni, 6) (unit in nm) is evaluated through ST-FMR measurements. A sign change in $\theta_{DL}$ is observed when NiO is around 2 nm. When $t_{NiO} \geq 3$ nm, the sign of $\sigma_M^{eff}$ changes from negative to positive, indicating the dominance of the orbital-induced magnon torque (positive sign) over the spin-induced magnon torque (negative sign). Moreover, $\sigma_M^{eff}$ exceeds $\sigma_O^{eff}$, confirming that the magnon torque induced by *L*-*M* conversion governs the switching. Our results clearly demonstrate the superiority of *L*-*M* conversion for magnetization control, which is possibly attributed to the higher efficiency of *L*-*M* conversion compared to *S*-*M* conversion and the larger orbital Hall conductivity.

Figure 5e displays the switching efficiencies ($\eta$) for devices composed of NM (W, Ta, or Ti, 5)/NiO (3)/Ti (2)/CoFeB (0.9) and the control device W (5)/Ti (2)/CoFeB (0.9) without NiO (unit in nm). All devices exhibit PMA. Devices with NiO exhibit higher $\eta$, which is due to the large magnon torque from *L*-*M* conversion. Notably, the Ti/NiO-based device has $\eta$ nearly three times higher than that of the control device relying solely on spin torque (also see Supplementary Fig. 11). While the present demonstration uses Ti and NiO as a proof of principle, we anticipate that a broader range of orbital materials and magnetic insulators could be explored soon.

We demonstrate efficient excitation of magnon currents by orbital currents via *L*-*M* conversion at the interface between the orbital material (Ti) and the magnetic insulator (NiO). The measured *L*-*M* conversion coefficient is as high as ~0.06. Using this mechanism, we achieve robust room-temperature perpendicular magnetization switching in a CoFeB layer driven by orbital-induced magnon torque. The switching efficiency $\eta$ is much higher than that in devices relying solely on spin or orbital torques. Our work overcomes the limitations of weak orbital torque in traditional orbital material-ferromagnet heterostructures, highlights the advantages of *L*-*M* conversion for magnetization control, and significantly broadens the material platform for designing energy-efficient orbital- and magnon-based devices with perpendicular magnetic anisotropy.

**Methods**

**Film growth and device fabrication.** Nonmagnetic materials (NMs, NMs = W, Ta, Ti)/NiO ($t_{NiO}$)/ferromagnets (FMs) multilayers are deposited on thermally oxidized Si wafers at room temperature using high-vacuum magnetron sputtering with a base pressure below $4\times10^{-9}$ Torr. For second-harmonic and magnetization-switching devices, the FM layer consists of Ti (2)/$Co_{20}Fe_{60}B_{20}$ (0.9)/MgO (2)/$SiO_2$ (4) (unit in nm) with perpendicular magnetic anisotropy (PMA). These multilayers are patterned into Hall bar or strip devices with a 10-µm-wide channel using ultraviolet maskless lithography (TuoTuo Technology, UV Litho-ACA) and argon-ion etching. For the ST-FMR devices, the FM layer comprises $Co_{20}Fe_{60}B_{20}$ (6)/$SiO_2$ (4) or Ni (6)/$SiO_2$ (4) (unit in nm), with in-plane magnetic anisotropy, and the multilayer is patterned into a strip serving as a signal channel. Finally, a coplanar waveguide of Ta (3)/Cu (100)/Pt (2) (unit in nm) is fabricated.

**Second-harmonic and current-induced magnetization switching measurements.** For second-harmonic measurements, an AC current at 133.33 Hz with an amplitude of 2 mA, generated by a Keithley 6221 current source, is applied to the Hall bar devices. The first- and second-harmonic Hall voltages are measured using SR830 lock-in amplifiers. The torque efficiencies are calculated from the effective field, with corrections for planar Hall effect (PHE) and anomalous Nernst effect (ANE) (see Supplementary Note 1). For the current-induced magnetization switching, two pulsed currents from a Keithley 6221 current source are applied to the devices. The first pulsed current, with a width of 50 μs, is used to drive perpendicular magnetization switching, followed by the second pulsed current, serving as the measurement current, with a magnitude of 2 mA and a width of 200 μs. Meanwhile, a Keithley 2182A nanovoltmeter records the anomalous Hall voltage. Temperature-dependent measurements are performed using a physical property measurement system (PPMS-DynaCool-9T, Quantum Design).

**ST-FMR measurements.** An in-plane RF current $I_{RF}$, with frequencies ranging from 6 GHz to 9 GHz and a power of 10 dBm, is applied at fixed angles ($\varphi_H$ = 38°) relative to the in-plane external magnetic field. The ST-FMR voltage, $V_{mix}$, is measured using an SR830 lock-in amplifier.

**MOKE imaging measurements.** For magneto-optical Kerr microscope (MOKE) imaging measurements, the CoFeB magnetization is first saturated along the +*z*-axis with an out-of-plane external magnetic field *H*. Then, we remove *H* and apply a pulsed current to induce the CoFeB magnetization switching. The magnetization states are then captured by the magneto-optical Kerr microscope (TuoTuo Technology, TTT-02-Kerr Microscope) after the pulsed current is turned off.

**Data availability**

The data that support the findings of this study are available in the paper and the Supplementary Information. Other relevant data are available from the corresponding authors upon reasonable request.

**Acknowledgements**

The work is supported by the National Natural Science Foundation of China (T2495211, 12261131506, 12074052). Guangdong Provincial Quantum Science Strategic Initiative (GDZX2401002). The Natural Science Foundation of Liaoning Province of China (2021-YQ-06). The Open Fund of the State Key Laboratory of Spintronics Devices and Technologies (SPL-2410). The Fundamental Research Funds for the Central Universities (DUT24GJ204, DUT25Z2746). The authors thank Dr. Jingyi Xiao from the Instrumental Analysis Center, Dalian University of Technology, for her assistance with electrical transport measurements.

**Author contributions**

F.M. and Y.W. conceived and designed the experiments. B.K. and F.M. grew the samples. Y.F., M.S. and D.S. performed ST-FMR measurements and device fabrications. M.S. and T.Z.

analysed the ST-FMR data. F.M. and Y.W. wrote the paper. J.Z., W.Z. and J.J.Z. provided discussion. J.J.Z., J.Z., F.M. and Y.W. analysed the data. Y.W. and J.J.Z. supervised the project. All authors discussed the results and commented on the paper.

**Competing interests**

The authors declare no competing interests.

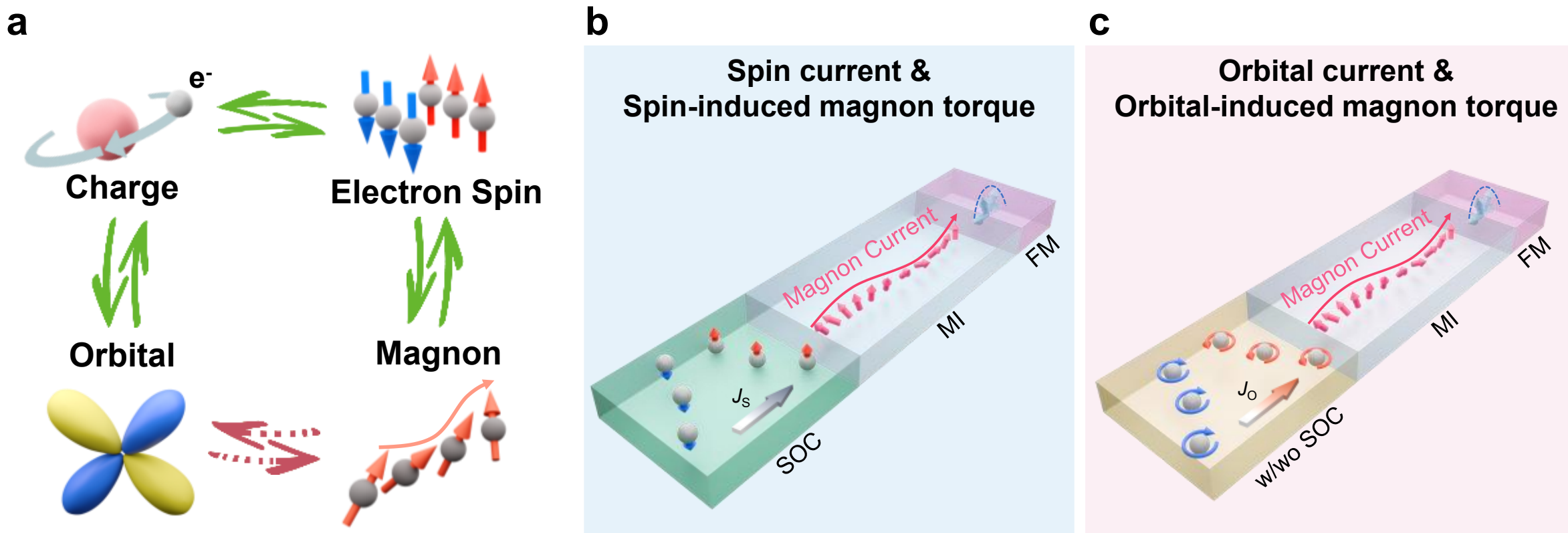

**Fig. 1 | Orbital angular momentum-to-magnon (*L*-*M*) conversion and orbit-induced magnon torque. a,** The conversion between different degrees of freedom. Here, the conversion between orbitals and magnons remains elusive. **b**, A schematic illustration showing the conversion between electron spins and magnons in magnetic insulators (MIs) and the spin-induced magnon torque effect. **c**, A schematic illustration depicting the conversion between orbitals and magnons in MIs and the orbital-induced magnon torque effect.

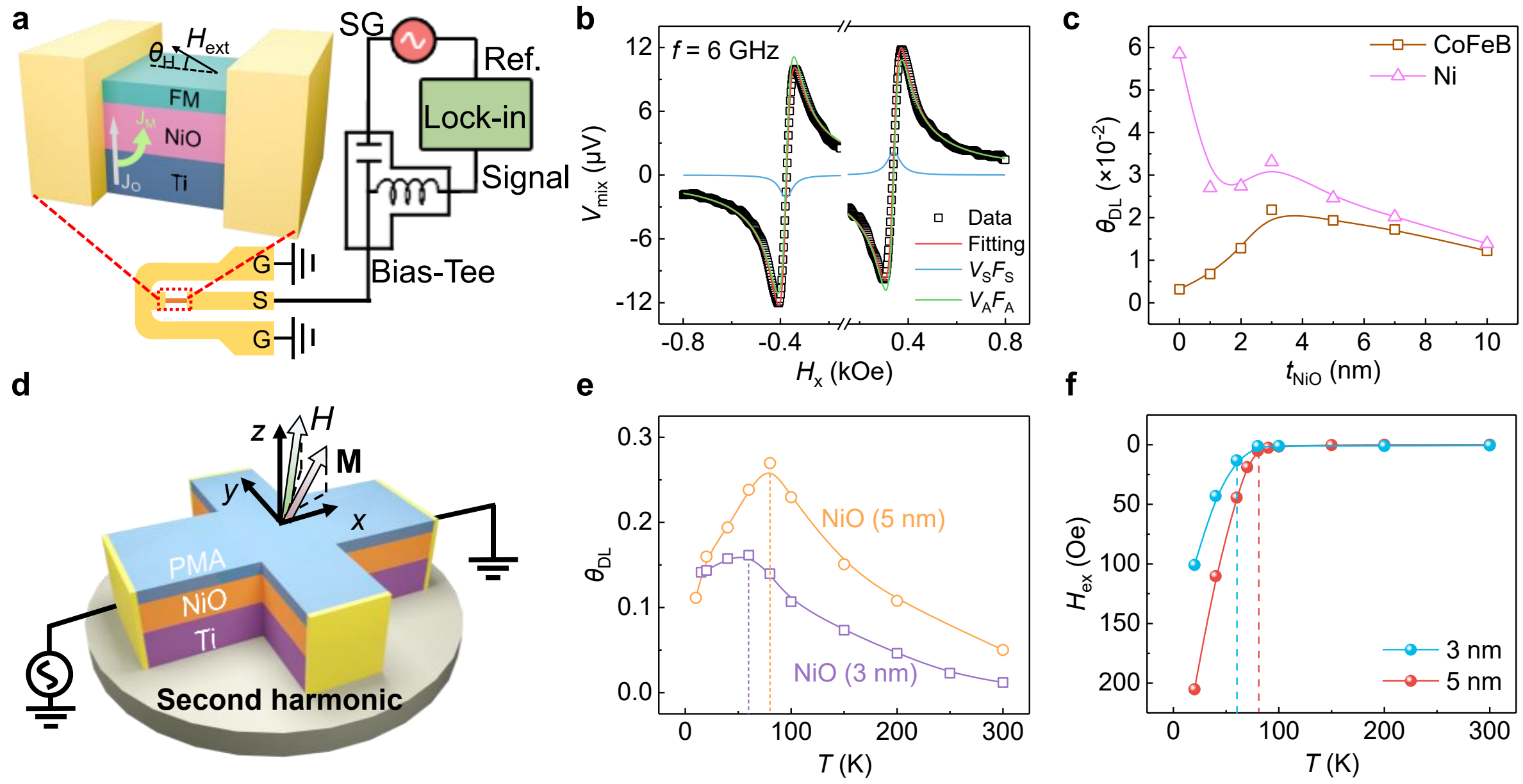

**Fig. 2 | Observation of orbital-induced magnon torque in Ti/NiO ($t_{NiO}$)/ferromagnets. a,** A schematic illustration of the ST-FMR measurements. The orbital current (gray arrow) and the magnon current (green arrow) are absorbed by the ferromagnetic layer (FM), and the resulting ST-FMR signal is detected by a lock-in amplifier through a Bias-Tee. **b**, A representative ST-FMR signal of the Ti (5 nm)/NiO (3 nm)/CoFeB device at 6 GHz and the fits from the symmetry ($V_S F_S$; blue) and anti-symmetry ($V_A F_A$; green) Lorentzian functions. **c**, NiO thickness ($t_{NiO}$) dependence of damping-like torque ($\theta_{DL}$) in Ti/NiO/CoFeB (brown) and Ti/NiO/Ni (pink). **d**, The Hall device geometry of Ti (5 nm)/NiO ($t_{NiO}$)/FM (Ti/CoFeB) with PMA and an illustration of second harmonic measurements. **e**, The $\theta_{DL}$ as a function of temperature from second-harmonic measurements for devices with $t_{NiO}$ = 3 nm (purple) and $t_{NiO}$ = 5 nm (orange). **f**, The exchange bias $H_{ex}$ as a function of temperature for Ti/NiO ($t_{NiO}$)/NiFe control devices at $t_{NiO}$ = 3 nm (sky blue) and $t_{NiO}$ = 5 nm (rose). The temperature at which the peak of $\theta_{DL}$ appears in (**e**) is consistent with the Néel temperature of NiO for the corresponding thickness.

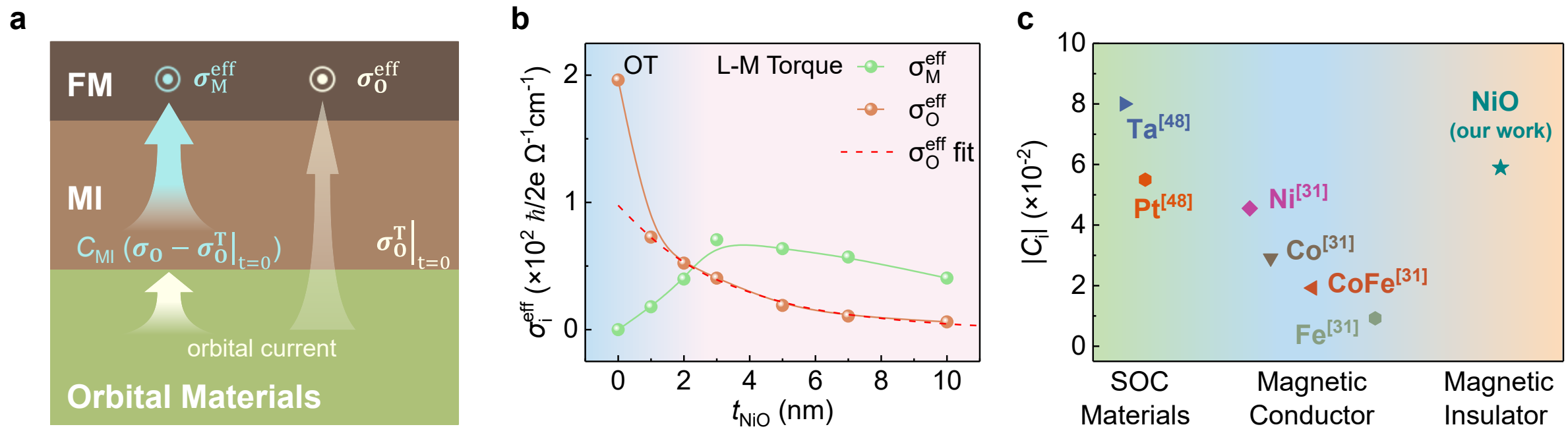

**Fig. 3 | Efficient *L-M* conversion in the magnetic insulator NiO. a**, Schematic illustration of magnon current (the medium arrow) converted from the orbital current (the small arrow) via the *L-M* conversion mechanism at the interface between orbital material and magnetic insulator, and the remaining orbital current transmission through the NiO layer (the large arrow). Both the orbital current and the magnon current can reach the FM layer and exert torques. **b**, Magnon-originated effective Hall conductivity ($\sigma_M^{eff}$) and orbital-originated effective Hall conductivity ($\sigma_O^{eff}$) converted from orbital current as functions of NiO thicknesses and the fit. **c**, Comparison of *L-M* conversion coefficient in the magnetic insulator NiO and *L-S* conversion coefficients in reported magnetic conductors (Ni, Co, CoFe and Fe)[31] and non-magnetic conductors with large spin-orbit coupling (SOC)[47].

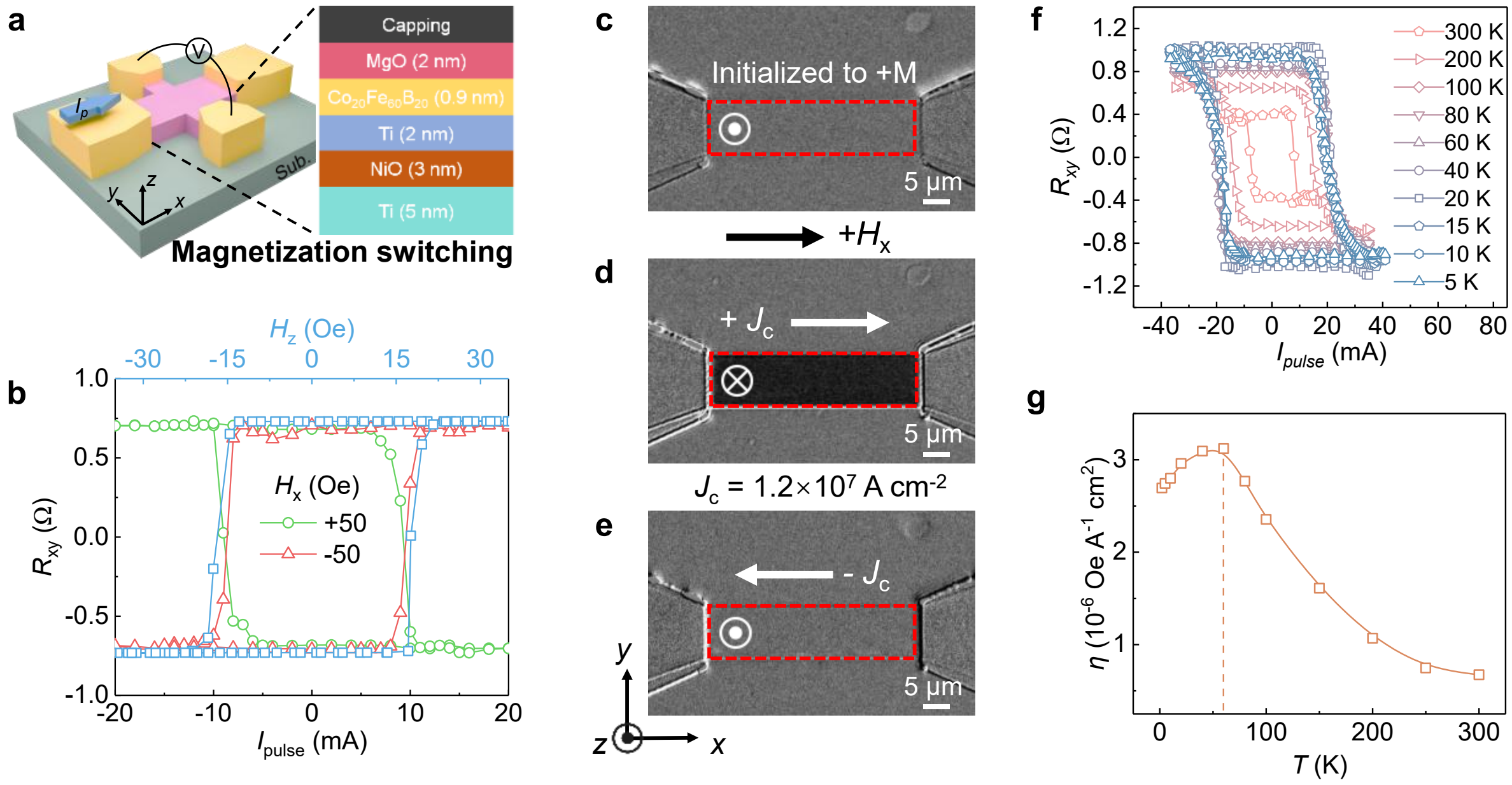

**Fig. 4 | *L-M* conversion-induced magnon torque-driven perpendicular magnetization switching in the Ti/NiO/Ti/CoFeB devices with PMA. a**, Schematic of the switching device and the multilayer structure. **b**, The variation of $R_{xy}$ driven by pulsed current (green and red) and out-of-plane magnetic field (blue). **c-e**, MOKE images of current-induced magnetization switching (denoted by the red rectangular box) by injecting a pulsed current $I_{pulse}$ parallel (**d**) and antiparallel (**e**) to the in-plane magnetic field $H_x$ at room temperature. The amplitudes of $I_{pulse}$ and $H_x$ are 11 mA and 50 Oe, respectively. The light (dark) contrast represents magnetization along the $+z$ ($-z$) axis. Before applying the pulsed current, the magnetization is initialized to the $+z$ direction. **f**, Pulsed current-induced magnetization switching with $H_x$ of +50 Oe at different temperatures ranging from 5 K to 300 K. **g**, Temperature dependence of the switching efficiency ($\eta$) derived from the data in (**f**).

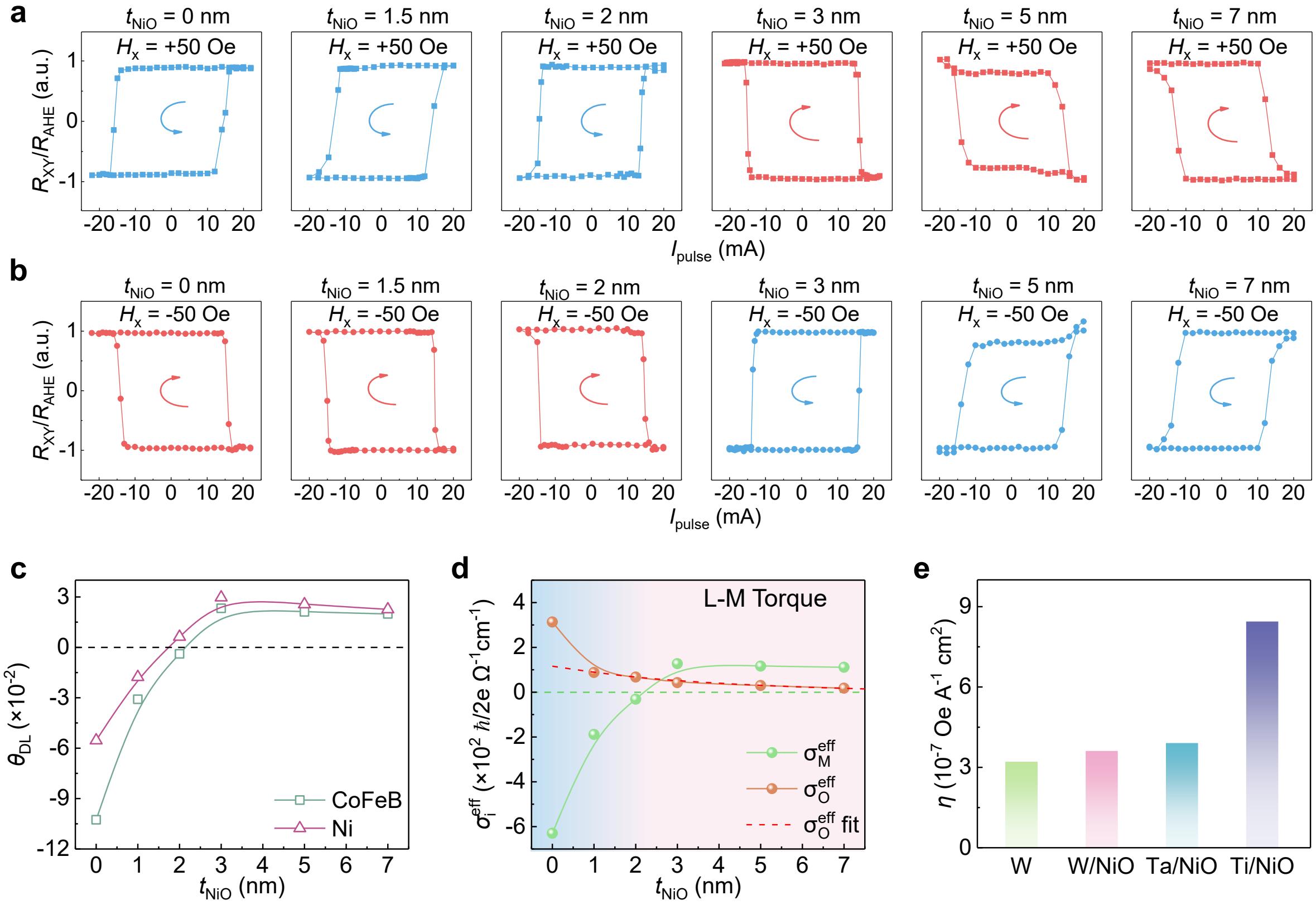

**Fig.5 | Superiority of *L-M* conversion in magnetization switching. a-b**, Polarity change of the current-induced magnetization switching in W (5 nm)/NiO ($t_{NiO}$ = 0, 1.5, 2, 3, 5, and 7 nm)/Ti/CoFeB devices with PMA under in-plane magnetic fields of +50 Oe in (**a**) and −50 Oe in (**b**). The blue and red arrows indicate the corresponding switching polarities. **c**, NiO thickness dependence of $\theta_{DL}$ in W (5 nm)/NiO ($t_{NiO}$)/CoFeB (6 nm) (olive) and W (5 nm)/NiO ($t_{NiO}$)/Ni (6 nm) (purple), separately. **d**, NiO thickness dependence of $\sigma_M^{eff}$ and $\sigma_O^{eff}$. The red dashed line represents the fit of $\sigma_O^{eff}$ against $t_{NiO}$. **e**, The switching efficiencies ($\eta$) for W (5 nm)/Ti (2 nm)/CoFeB (0.9 nm) and NM (W, Ta, or Ti, 5nm)/NiO (3 nm)/Ti (2 nm)/CoFeB (0.9 nm) with PMA.